\documentstyle[12pt]{article}

\begin{document}

\begin{titlepage}
\begin{flushright}
\par\vglue -2cm
CERN--TH/95-86  \\
ENSLAPP-A-514/95\\
ISN 95--28\\
ITKP 95--11
\end{flushright}
\vfill
\begin{center}
{\Large{\bf \protect{\boldmath$\Omega_{\rm c}$\unboldmath}}}\\
\vskip .3cm
{\Large{\bf  AND OTHER CHARMED BARYONS}}\\
\vskip .3cm
{\Large{\bf REVISITED}}\\
\vskip .6cm
{\bf Andr\'e Martin}\\
\vskip .1cm
{\small CERN, Theory Division}\\
{\small  CH--1211 Gen\`eve 23}\\
\vskip .1cm
{\small and}\\
\vskip .1cm
{\small  Laboratoire de Physique Th\'eorique ENSLAPP$^\star$}\\
{\small  Groupe d'Annecy, LAPP, B.P.\ 110}\\
{\small  F--74941 Annecy-le-Vieux, France}\\
\vskip .2cm
{\bf Jean-Marc Richard}$^\dagger$\\
\vskip .1cm
{\small Institut des Sciences Nucl\'eaires--CNRS--IN2P3}\\
{\small Universit\'e Joseph Fourier}\\
{\small 53, avenue des Martyrs, F-38026 Grenoble, France}\\
\vskip .1cm
{\small and}\\
\vskip .1cm
{\small Institut f\"ur Theoretische Kernphysik}\\
{\small Rheinische Friedrich-Wilhelms Universit\"at}\\
{\small Nu\ss allee 14--16, D--52115 Bonn, Germany}
\vglue 1.5cm
{\bf Abstract}\\
\vglue.2cm
\end{center}
The mass of the $\Omega_{\rm c}^0$ baryon with quark content (ssc) is
computed in a potential model whose parameters have been determined in
1981 by fitting the spectrum of heavy mesons. It is found in perfect
agreement with a recent measurement at the CERN hyperon-beam
experiment.  The spectroscopy of other charmed baryons in potential
models is briefly reviewed.\par \vfill
\begin{flushleft}
{\small ${}^\dagger$ Supported by a A.\ von
  Humboldt French-German Research Grant}\\
{\small ${}^\star$ URA 14-36 du CNRS associ\'ee \`a l'\'Ecole Normale
  Sup\'erieure de Lyon et \`a l'Universit\'e de Savoie}\\
CERN--TH/95-86\\
\today.
\end{flushleft}
\end{titlepage}

%
Many potential models have been successfully tuned to reproduce the
spectrum and static properties of heavy quarkonia. A key property is
{\sl flavour independence}\/: the same potential holds for various
quark--antiquark systems. Small corrections are however expected, in
particular in the spin-dependent part. They  arise when reducing any
relativistic kernel into the Schr\"odinger framework.

One of these models with flavour independence built in is the simple
power-law potential designed by one of us \cite{Martin81}
\begin{equation}
\label{eq:martin-central}
V=-8.064+6.8698\,r^{0.1},
\end{equation}
where units are GeV for the potential $V$, and GeV$^{-1}$ for the
interquark distance $r$. This is a variant of the logarithmic
potential which produces the same spacings for all Q$\overline{\rm
  Q}$' spectra \cite{Log}. The central potential
(\ref{eq:martin-central}) is supplemented by a spin--spin term of
contact type
\begin{equation}
\label{eq:martin-spin-spin}
V_{\rm SS}=1.112 \,{\vec\sigma_1\!\cdot\vec\sigma_2\over m_1m_2}
\delta^{(3)}(\vec{\rm r}),
\end{equation}
where the $m_i$ are the constituent masses. $V_{\rm SS}$ is treated at
first order, and is adjusted to reproduce the correct J$/\psi-\eta_{\rm
c}$
hyperfine splitting of charmonium. The quark masses are
\begin{equation}
\label{eq:martin-masses}
m_{\rm s}=0.518,\qquad  m_{\rm c}=1.8,\qquad m_{\rm b}=5.174\;{\rm
  GeV}.
\end{equation}
This model has been very successful in reproducing the masses of the
${\rm c\bar c}$, ${\rm b\bar b}$, ${\rm s\bar s}$ and ${\rm c\bar s}$
bound states and its predictions for the
${\rm b\bar s}$ system have been
checked experimentally \cite{Martin-Fubini}.

The model of Eqs.(1-3) was applied in Ref.\ \cite{JMR1} to compute the
mass of the $\Omega^-$, using the semi-empirical rule
\begin{equation}
\label{eq:one-half}
V_{\rm QQ}={1\over2} V_{{\rm Q}\overline{\rm Q}},
\end{equation}
which is discussed in \cite{JMR1}, and references therein.
The result, 1662 MeV (actually now 1666 MeV from an improved
computation of the hyperfine contribution),
comes very close to the experimental value
${\cal M}(\Omega^-)=1672\;$MeV.
New, and accurate determinations of the mass of the $\Omega_{\rm c}^0$
are expected, and preliminary results are already reported.  A value
\begin{equation}
\label{WA-89-result}
{\cal M}(\Omega_{\rm c})=2706.8\pm 1\;\hbox{MeV}
\end{equation}
was presented at the Rencontres de Moriond, in march 1995, by the WA89
collaboration \cite{WA89Moriond}, which uses the CERN hyperon beam
\cite{WA89}. This value has to be compared with $2740\pm20\!$ MeV by
the WA62 CERN experiment \cite{WA62-Omega-c}, $2719\pm7\pm2.5\!$ MeV
by ARGUS \cite{ARGUS-Omega-c}, and $2705.9\pm3.3\pm2.0\!$ MeV by the
E687 experiment at Fermilab \cite{E687-Omega-c}. For a review on
experimental and theoretical aspects of heavy baryons, see, e.g.,
\cite{Koerner-review}.

It seems interesting to repeat the calculation of
\cite{JMR1} for (ssc) configurations with spin 1/2 $(\Omega_{\rm
  c})$, and 3/2 $(\Omega_{\rm c}^*)$. We find
\begin{equation}
\label{eq:mass-ssc}
{\cal M}(\Omega_{\rm c})=2708,\qquad {\cal M}(\Omega_{\rm
c}^\star)=2760\;
\hbox{MeV},
\end{equation}
using an hyperspherical expansion up to a ``grand'' orbital momentum
$L=8$ \cite{JMRrep}. Hence, it is expected that $\Omega_{\rm c}^\star$
will decay into $\Omega_{\rm c}$ with emission of a photon of about
$52\;$MeV.

Other predictions from the literature are listed in Table \ref{Tab1}.
\begin{table}
\caption{\label{Tab1} Comparison of predictions for the masses of
\protect{$\Omega_{\rm c}$} and \protect{$\Omega_{\rm c}^*$} baryons
(in MeV).}
\begin{center}
\begin{tabular}{lcll}
Authors&Ref.&\phantom{123}$\Omega_{\rm c}$&\phantom{123}$\Omega_{\rm
c}^*$\\
\hline
Roncaglia et
al.&\protect{\cite{Lichtenberg94}}&$2710\pm30$&$2770\pm30$\\
Samuel et al.&\protect{\cite{Samuel86}}&$2717\pm25$&$2767\pm35$\\
Izatt et al.&\protect{\cite{Izatt82}}&$2610$&$2710$\\
Rho et al.&\protect{\cite{Rho90}}&$2786$&$2811$\\
de R{\'u}jula et al.&\protect{\cite{DGG}}&$2680$&$2720$\\
Maltmann et al.&\protect{\cite{Maltman80}}&$2730$&$2790$\\
Chan&\protect{\cite{Chan77}}&$2773$&$2811$\\
Richard et al.&\protect{\cite{RiTa2}}&$2664$&$2775$\\
Silvestre-Brac&\protect{\cite{BSBssc}}&$2675$&$2749$\\
\multicolumn{2}{c}{Present work}&$2708$&$2760$\\
\hline
\end{tabular}
\end{center}
\end{table}
The estimate of Ron\-ca\-glia et al.\ \cite{Lichtenberg94} does not
result from a specific model, but from a survey of the regularities
of the hadron spectrum in flavour space. Not surprisingly, it comes
very close to the (preliminary) experimental mass. Ref.\
\cite{Samuel86} is a lattice calculation, \cite{Izatt82} a bag model.
The others are potential models. The closest to the present one is
\cite{RiTa2}, where the same functional dependence as in Eqs.\
(\ref{eq:martin-central},\ref{eq:martin-spin-spin}) is used, but,
there,
it is
attempted to fit all baryons, even those with light quarks, and this
results
in a larger strength for the hyperfine correction.

On the same line as \cite{Lichtenberg94}, one can derive inequalities
which do not depend on the specific choice of the potential $V$, and
would in fact hold for any flavour-independent model. Examples were
already given for beautiful baryons \cite{MaRi} and for light flavour
\cite{JMRrep}.  We assume that the interquark potential in baryons and
quarkonia satisfy
\begin{equation}
\label{eq:ineg-pot-baryon-meson}
V_{\rm QQQ}(\vec{\rm r}_1,\vec{\rm r}_2,\vec{\rm r}_3)\ge{1\over2}
\sum_{i<j} V_{{\rm Q}\overline{\rm Q}}(\vec{\rm r}_i-\vec{\rm r}_j),
\end{equation}
as for instance with the prescription (\ref{eq:one-half}), or with the
string-motivated model \cite{Yshape}
\begin{equation}
\label{eq:string-pot}
V_{{\rm Q}\overline{\rm Q}}=\lambda r,\qquad
V_{\rm QQQ}=\lambda\min_J(d_1+d_2+d_3),
\end{equation}
where $d_i$ is the distance from the $i^{\rm th}$ quark to a junction
$J$ whose location is adjusted to minimize the potential.

{}From Eq.\ (\ref{eq:ineg-pot-baryon-meson}), one easily derives
 an inequality between spin-averaged ground-state masses \cite{JMRrep}
\begin{equation}
\label{eq:simple-M3-to-M2}
{\cal M}({\rm ssc})\ge{1\over2}{\cal M}({\rm s \bar s})+{\cal M}({\rm
  c\bar s}).
\end{equation}
This lower bound can be estimated near $2.5\;$GeV, and is rather
crude. It can be improved in two ways. First one introduces hyperfine
corrections, assuming a linear dependence of the Hamiltonian upon the
spin operators $\vec\sigma_i\!\cdot\!\vec\sigma_j$. As shown in
\cite{MaRi}, one gets
\begin{equation}
\label{eq:improved-M3-to-M2}
{\cal M}(\Omega_{\rm c})\ge{1\over2}{\cal M}(\phi) + {3\over4}{\cal
  M}(D_{\rm s})+ {1\over4}{\cal M}(D_{\rm s}^\star)\ .
\end{equation}
One can also use the Hall--Post techniques \cite{Post} to remove the
centre-of-mass energy left over in 2-body subsystems when deriving
(\ref{eq:simple-M3-to-M2}). Let us simplify the notations into $m_{\rm
s}=1$
and $m_{\rm c}=M$, with typically $M\simeq3-4$. The kinetic
energy of the baryon
\begin{equation}
\label{kinetic1}
T={\vec{\rm p}_1^2\over2}+{\vec{\rm p}_2^2\over2}+{\vec{\rm
    p}_3^2\over2M}
\end{equation}
can be rewritten as \cite{BMR3}
\begin{eqnarray}
\label{kinetic2}
&&T=\left(\vec{\rm p}_1+\vec{\rm p}_2+\vec{\rm p}_3\right)\cdot
\left(b\vec{\rm p}_1+b\vec{\rm p}_2+b'\vec{\rm p}_3\right)\nonumber\\
&&\ \ \ {}+\sum_{i=1,2}{a_{i3}\over2}\left({ \vec{\rm p}_i-x\vec{\rm
p}_3\over
  1+x}\right)^2 +{a_{12}\over2} \left({\vec{\rm p}_1-\vec{\rm
    p}_2\over2}\right)^2\ ,
\end{eqnarray}
where, for given $x$, the inverse masses are
\begin{eqnarray}
\label{eq:values-of-a}
&&a_{23}=a_{31}=\left({1+x\over1+2x}\right)^2\left(1+{2\over
  M}\right)\nonumber\\
&&a_{12}={8\over(1+2x)^2}\left(x(1+x)-{1\over2M}\right)\ .
\end{eqnarray}
This gives the following inequality on Hamiltonians
\begin{equation}
\label{H3-to-H2}
H_3=T+V_{\rm QQQ}\ge {1\over2}\sum_{i<j}H_2(a_{ij},V_{\rm Q\overline{\rm
Q}}),
\end{equation}
where $H_2(a,V)=-a\Delta+V$. This implies an inequality on
ground-state energies
\begin{equation}
\label{E3-to-E2}
E_3({\rm ssc})\ge\sum_{i<j}E_2[a_{ij}]\ .
\end{equation}
The l.h.c. can be optimized by varying $x$,  leading to
inverse masses $a_{ij}$ which are larger than the inverse
masses $b_{ij}=(m_i^{-1}+m_j^{-1})/2$ in actual ${\rm s}\bar{\rm s}$ or
${\rm
  c}\bar{\rm s}$ mesons. The corresponding change of binding energy
can be bounded in terms of the orbital excitation energy $\delta
E=E_2({\rm 1P})-E_2({\rm 1S})$. The result is \cite{BMR2}
\begin{equation}
\label{eq:change-of-E2}
E_2[a]-E_2[b]\ge {3\over4}\left({a-b\over a}\right)\delta E[b],
\end{equation}
with mild restrictions on the Q$\overline{\rm Q}$ potential, $\Delta
V>0$, and $V''<0$, i.e., a behaviour in between Coulomb and linear.
If one estimates $\delta E$ form the experimental information on
positive-parity ${\rm s\bar s}$ and ${\rm c\bar s}$ mesons (with some
uncertainty for estimating the spin-averaged masses), and varies the
parameter $x$, one ends with a lower limit
\def\gappeq{\mathrel{\rlap{\raise .5ex\hbox{$>$}}
                          {\lower .5ex\hbox{$\sim$}}}}
\begin{equation}
\label{lower-limit-ssc}
{\cal M}(\Omega_{\rm c})\gappeq 2.65\;{\rm GeV}\ .
\end{equation}

The main conclusion of this study is that the ``1/2'' rule works
surprisingly well for relating mesons to baryons. The $\Omega$ and
$\Omega_{\rm c}$ masses are well reproduced. The predictions for the
ground state with other
flavour combinations are
\begin{equation}
\label{ccs-ccc}
{\cal M}(\Omega_{\rm cc})=3.737,\qquad{\cal M}(\Omega_{\rm cc}^
\star)=3.797, \qquad {\cal M}(\Omega_{\rm ccc})=4.787\;{\rm GeV}.
\end{equation}
There are reasonable expectations that these states, ``the ultimate
goal of baryon spectroscopy'' \cite{ccc1,Savage} would be seen at LHC.

\subsection*{Acknowledgments} One of us (A.M.) would like to repeat
his thanks to Murray Gell-Mann for suggesting him to include strange
quarks in potential models, He would also like to thank Tran Thanh Van
for the hospitality at the Rencontres de Moriond.


\end{document}